\pdfoutput=1

\documentclass[11pt]{article}

\usepackage[]{ACL2023}

\usepackage{times}
\usepackage{latexsym}

\usepackage[T1]{fontenc}

\usepackage[utf8]{inputenc}

\usepackage{microtype}

\usepackage{inconsolata}
\usepackage{bm}
\AtBeginDocument{%
  \providecommand\BibTeX{{%
    \normalfont B\kern-0.5em{\scshape i\kern-0.25em b}\kern-0.8em\TeX}}}
\usepackage{booktabs} 
\usepackage{algorithm}
\usepackage{algorithmicx} 
\usepackage{graphics}
\usepackage{ifthen}
\usepackage{amsmath}
\usepackage{latexsym}
\usepackage{CJK}

\usepackage{ifthen}
\usepackage{caption}
\usepackage{enumerate}
\usepackage{amsmath}
\usepackage{caption}
\usepackage{enumerate}
\usepackage{graphicx}
\usepackage{float}
\usepackage{bookmark}
\usepackage{amsmath}
\usepackage{multirow}

\usepackage{algpseudocode}
\usepackage{amsmath}
\usepackage{balance}

\usepackage{subfig}
\usepackage{graphics}
\usepackage{caption}
\usepackage{enumerate}
\usepackage{amsmath}
\usepackage{caption}
\usepackage{verbatim}
\usepackage{graphicx}
\usepackage{float}
\usepackage{amsmath}
\usepackage{multirow}
\usepackage{booktabs}
\usepackage{booktabs}
\usepackage{amsmath}
\usepackage[hang,flushmargin]{footmisc}
\usepackage{caption}
\usepackage{bbold}
\usepackage{color}
\usepackage[normalem]{ulem}
\usepackage{svg}

\usepackage{booktabs} 
\usepackage{tabularx} 
\usepackage{graphicx} 

\svgsetup{
	inkscapepath=i/svg-inkscape/
}
\svgpath{{svg/}}
\newcommand\numberthis{\addtocounter{equation}{1}\tag{\theequation}}
\newsavebox\CBox

\usepackage{newtxtext,newtxmath}

\usepackage{libertine}  
\usepackage[T1]{fontenc}  
\newcommand{\libertinebold}[1]{{\fontfamily{LinuxLibertineT-OsF}\fontseries{b}\selectfont #1}}

\newcommand{\paratitle}[1]{\vspace{1ex}\noindent{\bf #1}}

\title{Ploutos: Towards interpretable stock movement prediction with financial large language model}

\author{Hanshuang Tong, Jun Li, Ning Wu, Ming Gong\thanks{*Corresponding author} , Dongmei Zhang, Qi Zhang \\
Microsoft, Beijing, China\\  \{hanstong, junli1, wuning, migon, dongmeiz, zhang.qi\}@microsoft.com}

\begin{document}
\maketitle

\begin{abstract}

Recent advancements in large language models (LLMs) have opened new pathways for many domains. However, the full potential of LLMs in financial investments remains largely untapped. There are two main challenges for typical deep learning-based methods for quantitative finance. First, they struggle to fuse textual and numerical information flexibly for stock movement prediction. Second, traditional methods lack clarity and interpretability, which impedes their application in scenarios where the justification for predictions is essential. To solve the above challenges, we propose Ploutos, a novel financial LLM framework that consists of PloutosGen and PloutosGPT. The PloutosGen contains multiple primary experts that can analyze different modal data, such as text and numbers, and provide quantitative strategies from different perspectives. Then PloutosGPT combines their insights and predictions and generates interpretable rationales. To generate accurate and faithful rationales, the training strategy of PloutosGPT leverage rearview-mirror prompting mechanism to guide GPT-4 to generate rationales, and a dynamic token weighting mechanism to finetune LLM by detecting and emphasizing key tokens in rationales. Extensive experiments show our framework outperforms the state-of-the-art methods on both prediction accuracy and interpretability. 

\end{abstract}

\section{INTRODUCTION}

The prospect of predicting stock market trends has continually intrigued both academics and investors for a long time, and it is still challenging due to the volatile nature of financial markets and multimodal features related to stock price~\cite{adam2016stock}, such as textual features (e.g., news, tweets) and time-series numerical features (e.g. price sequence, volumes sequence)~\cite{kohara1997stock,neal1998measures}. 
In practice, researchers focus more on predicting stock movement (whether the stock price movement will increase or decrease) rather than attempting to predict the precise stock price values, which is widely recognized as a inherently unpredictable task.~\cite{walczak2001empirical}.

\begin{figure*}[htb]
	\centering
	\includegraphics[width=\linewidth]{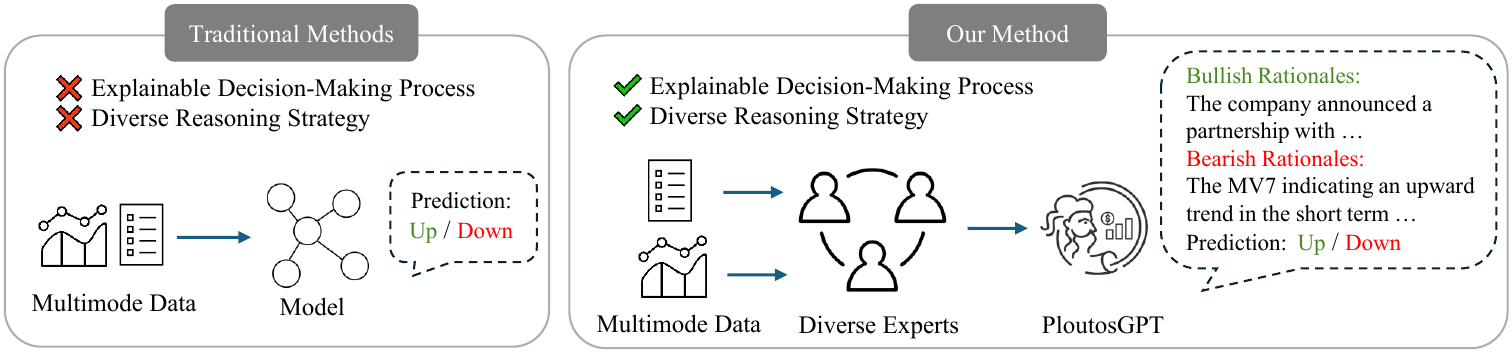}
	\caption{Comparison of Quantitative Methods}
	\label{fig:inspiration}
\end{figure*}

Quantitative methods can be either traditional deep model based or LLM based. Traditional methods use LSTM, transformer, or graph models to combine stock texts and prices~\cite{gao2023stockformer,sawhney2020deep}. However, they are opaque and lack of interpretability. LLM based methods use LLM to label sentiment or directly merge news and prices into model training~\cite{chen2023chatgpt,xie2023pixiu}. They also struggle to generate interpretable decision making rationales and fail to adaptively consider different strategies for different markets. To understand and fuse multi-modal features such as stock price and text data in a flexible way, we introduce PloutosGen pipeline that consists of multiple experts to model stock movement from different aspects.

\begin{enumerate}
    \item \textit{Technical Expert}, extracts technical features from time series based on various alpha formulas. Then it feeds these features into a time series forecasting model as next-token prediction. 
    \item \textit{Sentiment Expert}, captures the correlation between news events and their effect on stock valuations, which is well-established but heterogeneous and dependent on the content value of the news, ranging from breaking news to trivial commentary. This expert aggregates this diverse news with sophisticated methods.
    \item \textit{Human Expert}, incorporates human thoughts into the framework, which can enhance its effectiveness in real applications.
\end{enumerate}


However, the predictive accuracy of strategy experts can vary considerably across different stock situations. Therefore, we need an adaptive framework that can effectively incorporate the insights of the most suitable experts for any given stock situation. As shown in Fig.~\ref{fig:inspiration}, we leverage the interpretative and analytical capabilities of large language models to process multimodal data from diverse strategy experts, and enable them to adaptively learn and evolve their strategies to accommodate varying market conditions, and generate convincing decision making rationales. 

To cooperate with the PloutosGen pipeline, we further propose PloutosGPT. It consists of a rearview-mirror prompting strategy for data construction and a dynamic token weighting mechanism for supervised finetuning. Inspired by the common economic phenomenon that \textit{the rearview mirror is always clearer than the windshield}~\cite{Berkshire1991}, which indicates that analyzing the rationales of past cases is always easier than predicting the future, we prompt GPT4 to find suitable experts and use their strategies to account for the price variations each time there is a movement in stock prices. By this prompting strategy, we can collect high quality rationales for supervised finetuning of PloutosGPT. However, even with accurate and grounded rationales, we still face low accuracy in the key tokens generation of PloutosGPT. Hence, we propose a dynamic token weighting mechanism to detect and emphasize key tokens in supervised finetuning. It measures token importance by calculating cosine similarity between each token's hidden state and the type embedding produced by a verbalizer.

Our experiments demonstrate the effectiveness of our framework, which outperforms the state-of-the-art method on both prediction accuracy and interpretablity. To summarize, our contributions are as follow: 

\begin{itemize}


\item We propose PloutosGen, a novel financial LLM pipeline that analyzes and integrates multimodal data, particularly numeric time series data derived from various alpha formulas.

\item We propose PloutosGPT, an advanced financial LLM that uses rearview-mirror prompting for interpretable stock prediction and dynamic token weighting for supervised finetuning.

\item We introduce faithfulness and informativeness to evaluate the quality of generated rationales. Extensive experiments show the effectiveness and interpretability of our methodology.





\end{itemize}

\section{PROBLEM FORMULATION}
The goal of quantitative investment task is to learn stock price related information from both social media, specifically tweets, and past stock performance indicators to predict stock trends. The direction of a stock's movement is determined by comparing its adjusted closing prices over two consecutive trading days. This analysis is structured as a binary classification task, where the prediction revolves around whether the price of a particular stock will move up or down.

\subsection{Stock Prediction}

For a given stock $s \in \mathcal S$, a target prediction date $d$, and related historical prices and  news data, we define stock movement prediction label over a period of $\mathcal T$ days as:

$$Y_{d}=
\begin{cases}
\label{eq:p1}
0, & $$p_{d}^{c} < p_{d-1}^{c}$$\\
1, & $$p_{d}^{c} \geq p_{d-1}^{c}$$\\
\end{cases}$$

Here the $p_{d}^{c}$ represents the adjusted closing price in the target prediction date $d$~\cite{xu2018stock, sawhney2020deep}. The label $Y_{d}$ is categorized as $0$ if there's a decrease in the adjusted closing price, and as $1$ indicate a rise in the adjusted closing price.

\begin{figure*}[htb]
	\centering
	\includegraphics[width=\linewidth]{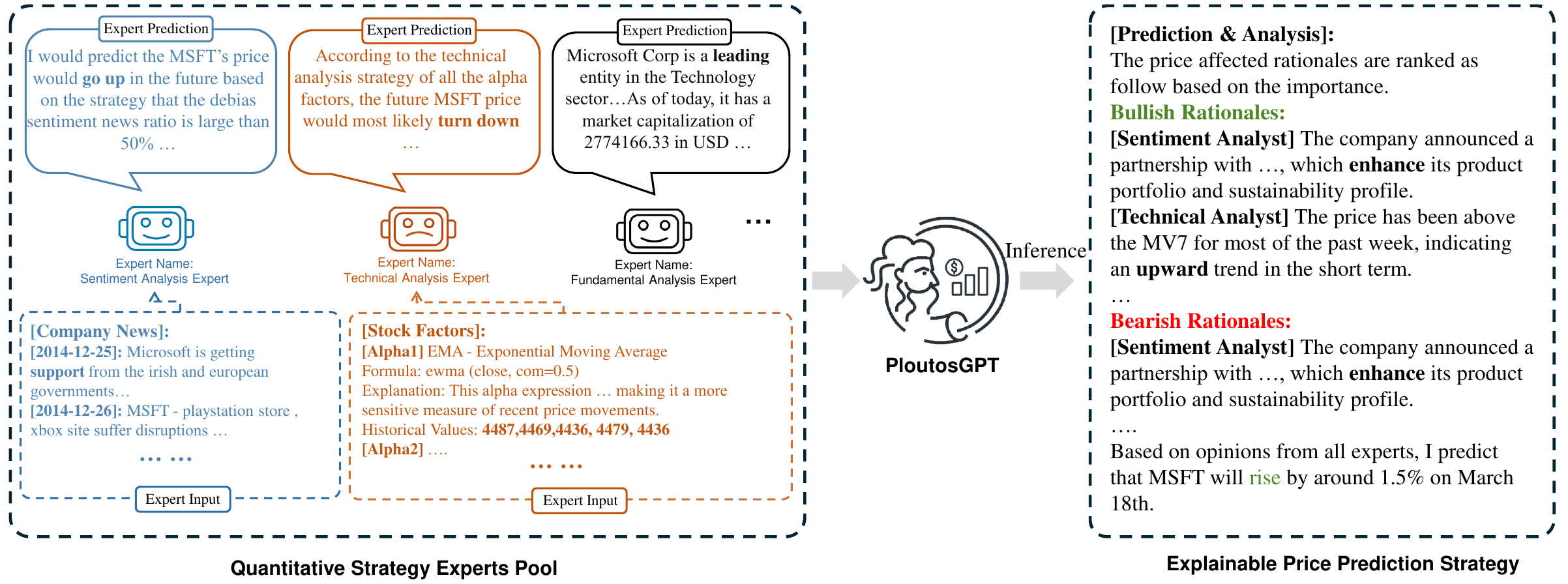}
	\caption{Ploutos framework: When predicting the price movement of a given stock, the diverse strategy experts pool would first leverage its diverse experts to gather and analysis the stock related feature with his strategy. Then, the PloutosGPT would generate bullish and bearish rationales to make final decision based on the market condition and opinions from diverse expert.}
	\label{fig:framework}
\end{figure*}

\section{METHODOLOGY}

In this section, we first give an overview of the framework of Ploutos, followed with a detailed explanation of PloutosGen and PloutosGPT.

\subsection{PloutosGen Pipeline}

As shown in Fig.~\ref{fig:framework}, the core component of PloutsGen pipeline is the diverse expert pool, which is designed to provide signals from a wide range of experts, including sentiment, technical analysis and human analysis, each utilizing a Large Language Model (LLM) to analyze the stock related information from a specific perspective. By integrating insights from these experts, the PloutosGPT can leverage a broad spectrum of knowledge and techniques to inform its predictive capabilities.




\subsubsection{Sentiment Analysis Expert}
In this section, we would introduce how we leverage only the sentiment information to predict the stock price. 
Previous research shows social media is useful for this task~\cite{si2013exploiting,si2014exploiting}, but there are two challenges: i) existing large language models (LLMs) are not fine-tuned for stock prediction from sentiment perspective; ii) how to aggregate news or tweets into stock signals, since different events affect stock price differently. We explore three ways based on whether stock signals are used for training: unsupervised, supervised, and supervised\&unsupervised sentiment-based stock prediction.

\paratitle{Unsupervised:} 
We use several datasets with different assets and news types, such as FIQA-SA~\cite{maia201818}, NWGI~\cite{chu2023data}, BTSA~\cite{kaggleBitcoinTweets}, and TNFS~\cite{zeroshottwitterfinancialnewssentiment}. These datasets ensure our models are trained on diverse financial information, enhancing robustness and adaptability. We use the sentiment instruction data to fine-tune LLMs~(see Table~\ref{tab:t2} in Appendix for detailed prompt), using LLaMA-2 as our backbone model. We evaluate the performance on ACL18 dataset~(see Appendix~\ref{sec:sent_data} for details). We use the trained sentiment model to label each news from the past 5 days ($\mathcal T=5$) of a stock as positive, negative, or neutral. The stock prediction on day $d$ is the average of sentiment scores. A positive score means an increase, while a negative score means a decrease. We combine all datasets to train the sentiment expert.

\paratitle{Supervised:} we construct the instruction with the prompt shown in Table \ref{tab:t3}, which takes the news text from the five days preceding the target day as input, and the stock price on the target day as the prediction label.
Similarly, we adopt LLaMA-2 as our base model for training and build our supervised method. We also try to combine the \emph{unsup} dataset to integrate both methods and improve the performance (more experiments are shown in Appendix~\ref{sec:sent_train}).




\begin{table}[ht]
	\centering
	\caption{Supervised Sentiment analysis prompt.}
	\label{tab:t3}
\resizebox{0.98\linewidth}{!}{%
\begin{tabular}{p{8cm}}
\bottomrule
\multicolumn{1}{c}{\textbf{Instruction Input}} \\
\hline
Instruction: Predict the MSFT's price movement for the 2015-06-10 based on the sentiment of the following tweets. \\
Please choose an answer from up or down.\\
Input: From 2015-06-05 to 2015-06-09, company related news during this period are listed below: \\
2015-06-08: game-changers announced at microsoft \\
2015-06-09: microsoft lumia sales tops apple in pakistan \\
2015-06-09: msft - are the days of subsidized 200 smartphones nearing an end? \\
Answer: \\
\hline
\multicolumn{1}{c}{\textbf{Instruction Output}} \\
\hline
The stock price of MSFT will go up.\\
\bottomrule
\end{tabular}%
}
\end{table}

\begin{table}[ht]
	\centering
	\caption{Technical analysis prompt. }
	\label{tab:tech}
 \resizebox{0.98\linewidth}{!}{%
\begin{tabular}{p{8cm}}
\bottomrule
\multicolumn{1}{c}{\textbf{Instruction Input}} \\
\hline
Instruction: Predict the MSFT's price movement for the 2015-06-10 based on the following tweets. \\
\textbf{[Stock Factors]}:\\
\textbf{[Alpha1]} MV7 - Moving Average of 7 Days \\
Formula: $\text{ts\_mean}(\text{close}, 7)$ \\
Explanation: This alpha expression ... which helps to smooth out short-term volatility and identify the underlying trend in the price movement. \\
Historical Values: 53, 67, 68, 84, 89 \\
\textbf{[Alpha2]} MV20 - Moving Average of 20 Days\\
…\\
Historical Values: 78,83, 92,87, 95\\
…\\
\textbf{[Analysis]}:\\
The historical price for MSFT from 2014-05-25 to 2014-05-28 is 63, 74, 78, 86. Please predict the MSFT's price for the 2014-05-29 based on all above information \\
Answer: \\
\hline
\multicolumn{1}{c}{\textbf{Instruction Output}} \\
\hline
The stock price of MSFT will be 93.\\
\bottomrule
\end{tabular}
}
\end{table}

\subsubsection{Technical Analysis Expert}
We aim to use large language models (LLMs) for multi-features time series forecasting in stock prediction, where we leverage multiple alphas (domain-specific time series features) and their explanations (text features) to frame the task as next-token prediction. However, existing methods are often limited by their number of time series features and deficiency in model explainability. To address these challenges, we propose Number-to-Text Alignment (N2I-Align), which uses prompt to integrate explaining texts with multi-features time series data. Table~\ref{tab:tech} illustrates our idea of using LLMs as a mediator between alphas and texts. We first generate multi-features time series datasets for a given stock, a target prediction date, and a fixed window size, using some common alphas (see Table \ref{tab:recent_basic_stock_factors} in Appendix for details). Then, we rescale and tokenize the time series data into sequences. To avoid the fragmentation and token capacity issues of conventional tokenization methods, we use the LLaMA-2 model, which naturally tokenizes long numbers into multiple digits. Finally, we use the N2I-Align prompt to wrap the tokenized sequences and generate the training data for technical expert tuning.

\subsubsection{Human Experts}
The human financial expert encapsulates the wisdom and intuition of seasoned investment professionals, including insights from fundamental analysis, macroeconomic indicators, and market cycles. The key feature of this expert is its ability to contextualize financial data within a broader economic and financial framework, providing an additional layer of natural language analysis that is often difficult to quantify but critical for making informed investment decisions. This expert serve as a customizable and optional component in PromptGen framework since it's not available in open-source dataeset but useful in real applications.

\subsection{PloutosGPT Training}

\begin{figure}[htb]
	\centering
	\includegraphics[width=\linewidth]{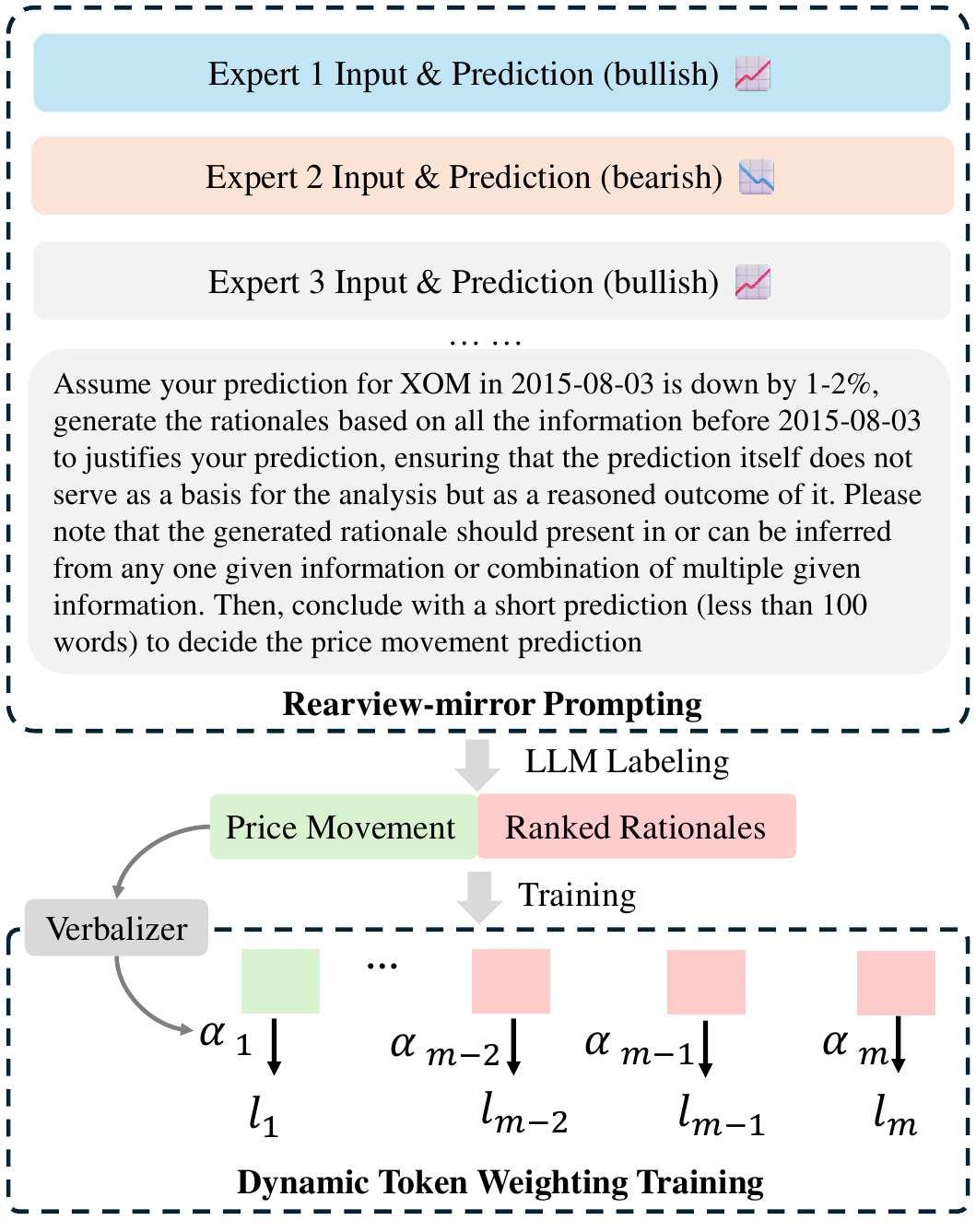}
	\caption{PloutosGPT is trained by dynamic token weighting and rearview-mirror data.}
	\label{fig:p_gpt}
\end{figure}
The main objective of PloutosGPT is to develop a Large Language Model (LLM) capable of generating explainable and accurate rationales for predicting price movements. Explainable AI in finance research has attracted increasing attention from various perspectives~\cite{weber2023applications}. For instance, some approaches aim to build interpretable stock price predictors using neuro-fuzzy rules~\cite{rajab2019interpretable,vlasenko2019novel}. Similarly, Yang et al.~\cite{yang2018explainable} leverage attention-based neural networks to assign importance to different news. However, the interpretability of these existing methods is largely compromised by the neural network learning phase. To the best of our knowledge, PloutosGPT is the first method that can generate natural language rationales for interpretable stock movement prediction. As shown in Figure~\ref{fig:p_gpt}, it consists of two steps: rearview-mirror based data construction and dynamic token weighting training strategy. In the first step, as shown in the figure, we prompt a powerful frozen LLM (e.g. GPT-4) to summarize both the bullish and bearish rationales for the price movement based on diverse experts' input, and generate rationales that is faithful to the given information (Since the ground truth is given, the frozen LLM knows which rationale makes the correct prediction). Then, the model is prompted to make the final decision based on the trade-off between bullish and bearish rationales. The intuition behind these prompts is to generate faithful rationales by learning weights for each expert's input based on their accuracy and relevance in different market conditions. All of these prompts are under the constraint that the model should not directly use the ground truth as the rationales, which echoes our initial motivation in introduction that researchers are better at analyzing the past given the result than predicting the future. 

To further finetune PloutosGPT, the generated rationales and stock movement label are combined as the output of the instruction training data, and the input and analysis from diverse experts are combined as the input. An example of the constructed instruction data is shown in the Table~\ref{tab:prompt_rearview} of Appendix.
By leveraging this curated training data, PloutosGPT becomes not just a black-box predictor but a sophisticated decision support tool that integrates the strengths of diverse strategies and provides actionable insights with a clear justification for its predictions. Moreover, though we have constructed bullish and bearish rationales based on ground truth and diverse experts' output, the fine-tuned PloutosGPT still struggles to generate key tokens of rationales. For example, in \emph{the MV7 indicating an upward trend in short term}, the key word \emph{upward} has to be treated carefully. Hence, we propose to dynamically control the weight of each token in the rationales of the training data as follows:




\begin{equation}
\numberthis 
\label{eq:f1}
\mathcal{L} = \sum_{i}  -\alpha_i \log{p(y_i|x, y_{1...i})}
\end{equation}
where $x$ is input of model, and $y$ is generated content consists of rationales and label. $\alpha_i$ is the weight of $i$-th token in generated content. It can be obtained by computing the similarity between a specific type embeddings and token embeddings:

\begin{equation}
\numberthis 
\label{eq:f2}
\alpha_i = \frac{\exp(\cos(\bm{t}_{Y_{d}}, \bm{h}_i)/\tau)}{\sum_{j=1}^m \exp(\cos(\bm{t}_{Y_{d}}, \bm{h}_{j})/\tau)}
\end{equation}
where $\bm{h}_i$ denotes hidden state of $i$-th token from transformer's last layers, $\bm{t}_{Y_{d}}$ is the type emebdding which is computed by average pooling of verbalizer tokens embedding and $\tau$ is temperature which controls the weight of key tokens. A suitable temperature could give more importance to key tokens, and hence bring higher prediction accuracy and high quality rationales. The verbalizer tokens are generated by verbalizer based on label $Y_{d}$. More details about the verbalizer can be checked in the Appendix~\ref{sec:verb}.

\section{EXPERIMENTS}
Our experiments are designed to investigate the following research questions:

\begin{itemize}  
  \item \libertinebold{RQ}\textbf{1}: How does Ploutos perform compared with current LLMbased and traditional prediction models? 
  \item \libertinebold{RQ}\textbf{2}: How do the different components in Ploutos affect its effectiveness?  
  \item \libertinebold{RQ}\textbf{3}: Does the decision making rationales informative and consistent with given information?
\end{itemize}

\subsection{Experiment Settings}

\subsubsection{Data}

In this study, we evaluate the proposed method on two public and classic datasets on stock movement prediction, ACL18~\cite{xu2018stock} and CIKM18~\cite{wu2018hybrid}. ACL18 comprises 88 high-volume trading stocks from the S\&P 500 index listed on the NYSE and NASDAQ exchanges. Each of stock contains price data obtained from Yahoo Finance and related tweets data extracted using regular expression queries based on NASDAQ ticker symbols for each day (except for weekends and public holidays). Then a $\mathcal T$-day ($\mathcal T$=5) rolling window is utilized to produce data candidates along the series of trading days. Samples are labeled as either positive or negative based on the percentage change in closing price, 
with increases of 0.55\% or more labeled as positive and decreases of 0.5\% or more labeled as negative, in order to generate an equitably partitioned dataset of 26,614 samples. Following Xu and Chen~\cite{xu2018stock}, we split the dataset into distinct sets for training (01/01/2014 to 31/07/2015), validation (01/08/2015 to 30/09/2015), and testing (01/10/2015 to 01/01/2016) in a 7:1:2 ratio. CIKM18 consists of 47 stocks with sufficient tweets and price data ranging from January 2017 to November 2017. Similarly, we leverage the same data processing method as ACL18 to process CIKM18 dataset, and we follow the same approach as stated in the original paper to split the datasets.

\subsubsection{Evaluation Metric}
Since Ploutos aims to predict price movement for a given target day, $i.e.,$ a binary classification problem. We adopt accuracy, F1 score, and the Matthews Correlation Coefficient (MCC, sourced from sklearn's implementations\footnote{https://scikit-learn.org}) for classification performance. We employ MCC due to its robustness and unbias evaluation of classifier performance regardless of data imbalance as it considers all quadrants of the confusion matrix.


\subsubsection{Baselines}
We compare Ploutos with the below baselines containing both traditional model based and LLM based methods. Those baselines are selected based on the criteria that it's either open-source or reproducible.

\noindent{$\textbf{Traditional Methods}$}: These methods uses traditional deep neural network such as time-series, graph based and others. Including ARIMA~\cite{brown2004smoothing}, Adv-LSTM~\cite{selvin2017stock}, StockNet~\cite{xu2018stock}, DTML~\cite{nguyen2015topic}.

\noindent{$\textbf{LLM Based Methods}$}: These methods leverage either zero shot or finetuned version of large language model to predict the stock movement. We select LLaMA-2~\cite{touvron2023llama}, GPT-4~\cite{achiam2023gpt}, FinMA~\cite{xie2023pixiu} as baselines.

\begin{table}[]
\centering
\caption{A comparison of prediction accuracy and MCC between Ploutos and other baselines.}
\label{tab:t31}
\resizebox{0.45\textwidth}{!}{
\begin{tabular}{ccccc}
\bottomrule[1.3pt]
\multicolumn{1}{c}{\multirow{2}{*}{\textbf{Models}}} & \multicolumn{2}{c}{\textbf{ACL18}} & \multicolumn{2}{c}{\textbf{CIKM18}} \\ \cline{2-5} 
\multicolumn{1}{c}{}      & \textbf{Acc.}  &\textbf{ MCC}   & \textbf{Acc.}   & \textbf{MCC}   \\ \bottomrule[1.3pt]
\textbf{ARIMA}         & 51.42 & -0.021 & 52.36 & -0.012      \\
\textbf{Adv-LSTM}         & 57.24 & 0.148 & 56.48 & 0.016      \\
\textbf{StockNet}      & 58.23 & 0.081 & 56.37 & 0.023      \\
\textbf{DTML}          & 57.44 & 0.191 & 58.62 & 0.045      \\ 
\textbf{GPT-4}          & 53.08 & 0.023 & 57.44 & 0.034      \\
\textbf{LLaMA-2 - 7B}        & 52.74 & 0.051 & 56.92 & 0.027      \\
\textbf{FinMA - 7B}     & 56.28 & 0.104 & 53.24 & -0.031      \\ \hline
\textbf{Ploutos - 7B}         & \textbf{61.21} & \textbf{0.205} & \textbf{59.89} & \textbf{0.064}      \\ \bottomrule[1.3pt]
\end{tabular}
}
\end{table}

\subsubsection{Training settings}
We conduct all experiments on 8 Tesla V100 GPU. To get the best performance of Ploutos, we apply grid search for batch size in \{64, 128, 256\} and learning rate in \{1e-6, 2e-06, 3e-6\} for each expert's training. Each experiments run three times and we report the average metric. Adam is used as the optimizer and we leverage our framework to fine-tune from LLaMA-2 for 2 epochs on all datasets. During the rearview-mirror data generation stage, we execute the scraping process using Azure OpenAI GPT-4 API. We use a temperature of 1.0 to generate samples and set the maximum number of tokens for generation to 2000. Moreover, we set the frequency penalty to zero and top-p to 1.0. For evaluation, we set the temperature to zero for all models to reduce output randomness and ensure more focused and deterministic outputs. 



\subsection{Performance Comparison (\libertinebold{RQ}\textbf{1})}

We compare the stock movement prediction performance of various methods including both traditional and LLM based methods. The evaluation results against baseline methods are presented in Table \ref{tab:t31}. From the figure and table, we draw the following conclusions: 1) Our method significantly outperforms both traditional and LLM based methods, verifying the superiority of tuning LLMs via our proposed training framework. Ploutos successfully unlocks the text and numerical data understanding capabilities of LLMs for stock movement prediction task. 2) Traditional deep models such as ARIMA, Adv-LSTM and others perform worse than Ploutos, implying the potential of leveraging language model to process and understand multi-mode stock related data. 3) The zero shot ability of LLaMA-2 and GPT-4 perform similarly to random guessing (ACC$\simeq 0.5$). However, our Ploutos achieves significant improvements. These results demonstrate a considerable gap between stock movement prediction and language tasks, showing the importance of using stock related data for tuning on LLMs. 4) Financial large language model FinMA-7B trained with StockNet training set yield inferior performance than Ploutos, indicating purely tuning LLM with stock related instruction data cannot get optimal performance for stock movement prediction task.

\begin{table}[]
\centering
\caption{Performance of Ploutos variants on ACL18 dataset, showing every component makes an important contribution to the excellent performance.}
\label{tab:t16}
\resizebox{0.39\textwidth}{!}{
\begin{tabular}{cccc}
\bottomrule[1pt]
\textbf{Model}    & \textbf{Accuracy} & \textbf{MCC} & \textbf{F1}    \\
\hline
Ploutos$_{\neg \text{S}}$        & 59.62    & 0.189 & 0.598  \\
Ploutos$_{\neg \text{T}}$         & 58.11    & 0.153 & 0.537  \\
Ploutos$_{\neg \text{R}}$          & 60.41    & 0.191 & 0.604 \\
\hline
\textbf{Ploutos}          & \textbf{61.21}    & \textbf{0.205} & \textbf{0.612} \\

\bottomrule[1pt]
\end{tabular}}
\end{table}

\subsection{Ablation Study (\libertinebold{RQ}\textbf{2})}

To evaluate the contribution of different components, we compare against several variants: 
\begin{itemize}
    \item \textbf{Ploutos$_{\neg \text{S}}$}: Ploutos without the sentiment analysis expert, which makes decisions just based on the numerical technical analysis. 
    \item \textbf{Ploutos$_{\neg \text{T}}$}: Ploutos without the technical analysis expert, which makes decisions just based on the news sentiment analysis. 
    \item \textbf{Ploutos$_{\neg \text{R}}$}: Ploutos without the rearview-mirror instruction data and dynamic token weighting mechanism.
    
\end{itemize}
The results are shown in Table \ref{tab:t16}. Ploutos$_{\neg \text{S}}$ achieves higher prediction accuracy than technical feature only methods such as ALSTM and Adv-LSTM, indicating the importance of the number to text alignment technique. Similar to the performance of Stocknet, Ploutos$_{\neg \text{T}}$ has a relatively high Acc. but lower MCC, suggesting texts are particularly helpful to predict on one side of the binary classification. By modeling both text data and stock relationships, Ploutos$_{\neg \text{R}}$ reaches a relatively good result. Finally, better performance achieved when rearview-mirror instruction data and dynamic token weighting are introduced into training. In summary, the ablation studies conclusively demonstrate that each component within the Ploutos architecture is important, and as a result the full model with all components achieves the best performance.      

\subsection{Faithfulness and Informativeness of Interpretablity (\libertinebold{RQ}\textbf{3})}

\begin{table}[]
\centering
\caption{Faithfulness and Informativeness of different methods on ACL18 dataset}

\label{tab:t19}
\resizebox{0.48\textwidth}{!}{
\begin{tabular}{cccc}
\bottomrule[1pt]
\textbf{Model}    & \textbf{Size} & \textbf{Faithfulness} & \textbf{Informativeness}     \\
\hline
FinMA         & 7B    & 59.53   & 61.32\\
FinGPT       & 7B    & 72.82   & 84.76\\
GPT-3        & 175B    & 77.63  & 91.58\\
\hline
\textbf{Ploutos}          & 7B    & \textbf{81.24} & \textbf{96.52}\\

\bottomrule[1pt]
\end{tabular}
}
\end{table}

To verify the quality of the interpretablity of the genareted rationales, we employ two quantifiable metrics: faithfulness and informativeness. Faithfulness measures whether the facts in model's response are based on or can be inferred from the given knowledge. Following the method in HaluEval~\cite{li2023halueval}, we design prompt to call GPT-4 to classify each fact within the model's response rationales as either inferable from the given knowledge or not, and then aggregate these accuracies as the model's faithfulness score (see Appendix \ref{sec:faithful} for more information). Informativeness measures the amount of information contained in the model's response, which is equally an important metric for interpretablity, as we found some models might output generally correct but uninformative rationales such as "\textit{predict rise in price due to the promising information.}". To measure the informativeness of each rationale, we directly use the informativeness measurement from TruthfulQA~\cite{lin2021truthfulqa}. 

Table \ref{tab:t19} shows the interpretablity score of different models, from which we can draw several conclusions: 1) Compared to other LLM methods, Ploutos demonstrates superior faithfulness and informativeness, which indicates that large language model can be trained through specific methods to generate informative and faithful rationales, thereby enhancing model's interpretability. 2) The FinMA-7B, which adopt similar training datasets compared to Ploutos, struggled to generate informative rationales, which indicate the effectiveness of the rearview-mirror data training strategy.  3) FinGPT~\cite{yang2023fingpt} outperforms FinMA in terms of both faithfulness and informativeness, which can be attributed to its chain of thought scraping and training technique to enhance reasoning capabilities. However, compared to Ploutos, it does not take technical indicators such as the interpretability of various alphas into account, leading to a lack of technical analytical capabilities.

\begin{figure}[!ht]
	\centering
         \hspace*{-0.3cm}
	\vspace{-5pt}
         \includegraphics[width=0.45\textwidth]{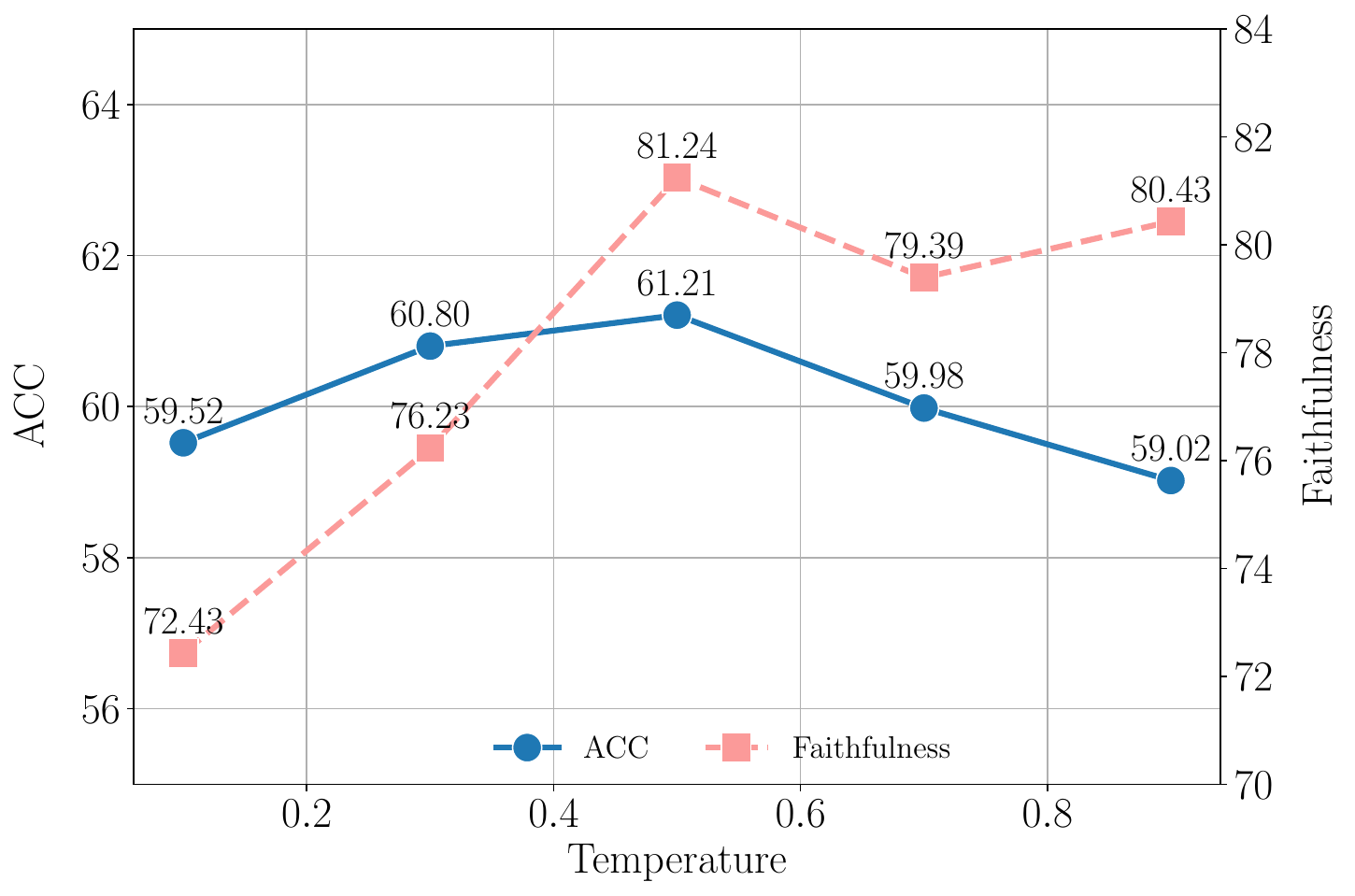}

	\caption{ACC and faithfulness over different temperatures on ACL18 Dataset}
	\label{fig:temp}
	\vspace{-10pt}
\end{figure}

Moreover, we explored the impact of temperature of dynamic token weight on the model's predictive accuracy and interpretability. The results are summarized in Fig.~\ref{fig:temp}. It can be observed that when the temperature is greater than 0.5, a lower temperature corresponds with higher accuracy, aligning with our intuition that a lower temperature increases the weight of the loss focus on predicting the stock movement.  However, when the temperature is less than 0.5, both accuracy and faithfulness begin to decrease, indicating that overly focusing on the movement itself and neglecting the interpretable rationale tokens can actually weaken the model's reasoning capabilities. When temperature equals to 0.5, both faithfulness and accuracy reach their maximum, suggesting a synergistic rather than a mutually exclusive relationship between predictive accuracy and interpretability, which also demonstrates the effectiveness of dynamic token weighting.

\section{Related Work}\label{2}

\subsection{Stock Prediction with Deep Learning} Deep learning approaches have gained prominence for their capacity to model complex patterns in stock market data. It can be divided into several tracks: RNN Based, CNN Based, Transformer Based and Graph Based Models. Recurrent neural networks (RNNs) are favored for their proficiency in capturing sequential trends in stocks~\cite{li2018stock,qin2017dual}, with advancements like Long Short-Term Memory (LSTM) networks enhanced by adversarial training demonstrating superior performance in stock prediction~\cite{feng2018enhancing}. Moreover, CNN Based Models incorporating stock times series data as a set of input feature maps~\cite{hoseinzade2019cnnpred,lu2021cnn}. Furthermore, the emergence of transformer-based models, exemplified by StockFormers~\cite{gao2023stockformer}, showcases the effectiveness to extract latent representations of long short term stock states and asset relations. Recent efforts have also explored graph-based deep learning models or casual matrix to analyze the interconnections among stocks, outperforming price-only models~\cite{luo2023causality, sawhney2020spatiotemporal,kim2019hats}.


\subsection{Stock Prediction with Large Language Model} 
The recent strides in large language model (LLM) have catalyzed the exploration of stock prediction application with LLM. Existing studies consists of several directions. Some studies focus solely on textual information. Lopez-Lira et al.~\cite{lopez2023can} demonstrate the correlation bewteen ChatGPT sentiment score and subsequent stock movement. Some researchers focus more on 
the numerical feature, Alpha-GPT is proposed as a new alpha mining paradigm by leveraging the power of large language models~\cite{wang2023alpha}. Others propose methods combine numerical and textual features, Xie et al.~\cite{xie2023pixiu,xie2023wall} directly prompt and fine-tune LLM to leverage both news and numerical features. Futhormore, Chen et al.~\cite{chen2023chatgpt} explore the relationship between stock by analyzing the textual information by ChatGPT. However, these techniques fall short of exploring the interpretablity ability of LLM, thereby limiting their application potential.


\section{CONCLUSION}

In this work, we explored the feasibility of using LLMs to construct a framework that can generate interpretable decision making rationals for stock movement prediction. Specifically, we first explore different strategies to analyze the multi-model stock related data for stock movement prediction from different perspectives. Then we propose PloutosGPT, which introduces an novel training approach called rearview-mirror prompting and dynamic token weightin that are designed to dynamically incorporate insights from diverse quantitative strategy experts, encompassing both sentiment and technical analysis. Extensive experiments shows that our method can perform well under different conditions, reinforcing its practical applicability.





\section*{Limitations}
While our proposed Ploutos framework shows promising results in stock movement prediction, there are several limitations that need to be addressed:
\begin{itemize}
    \item The PloutosGPT is designed to combine insights and predictions from multiple primary experts. However, the selection and performance of these experts can significantly impact the model's performance. The process of expert selection and their potential biases need to be thoroughly examined.
    \item The computational cost of our model, especially when dealing with large-scale data, can be high. Optimizing the model for efficiency without compromising its predictive accuracy is a challenge.
    \item Our model currently focuses on textual and numerical data. Incorporating other types of data, such as visual data, could potentially improve the model's performance.
\end{itemize} 

\bibliography{ploutosgpt}
\bibliographystyle{acl_natbib}

\appendix


\section{Sentiment Experts}

\subsection{Choice of unsupervised dataset}
\label{sec:sent_data}
Table~\ref{tab:t1} shows the comparison of the impact for stock movement prediction task over different datasets. Among all the options, the \emph{unsupervised} method combining all the datasets performs the best, which is reasonable since this method consists of most diversified and robustness data for sentiment analysis. Besides,  the accuracy of stock market predictions using sentiment scores calculated by this unsupervised method hovers around 50\%, which is not particularly high. This indirectly suggests that while the sentiment of these news articles does have a positive correlation with stock prices, there is also a lot of noise within the news itself. We need to develop better methods to aggregate this news.

\begin{table}[ht]
	\centering
	\caption{Comparison of the impact for stock movement prediction task over different datasets. The definition of MCC are introduced in section A. The Unsup. method denote that the combined dataset including both FIQA-SA, NWGI, BTSA, TNFS datasets. }
	\label{tab:t1}
  \resizebox{0.98\linewidth}{!}{%
\begin{tabular}{lllll}
\bottomrule
\textbf{Dataset}                & \textbf{Accuracy}          & \textbf{MCC}           &  &  \\
\hline
FIQA-SA~\cite{maia201818}                & 52.92          & 0.032          &  &  \\
NWGI~\cite{chu2023data}                   & 52.48          & 0.014          &  &  \\
BTSA~\cite{kaggleBitcoinTweets}                   & 52.56          & 0.043          &  &  \\
TNFS~\cite{zeroshottwitterfinancialnewssentiment}                   & 51.12          & -0.021         &  &  \\
\hline
Unsup. & \textbf{53.24} & \textbf{0.052} &  &  \\
\bottomrule
\end{tabular}
}
\end{table}

\subsection{Comparison of different training methods}
\label{sec:sent_train}
 Table \ref{tab:t4} shows the prediction results on StockNet for different methods. The results indicate that the combined 'sup \& unsup' method performs the best, which also demonstrates the effectiveness of both our supervised and unsupervised approaches.

\begin{table}[ht]
	\centering
	\caption{Comparison of the impact for stock movement prediction task over different sentiment analysis methods. }
	\label{tab:t4}
\begin{tabular}{lllll}
\bottomrule
\textbf{Method}                & \textbf{Accuracy}          & \textbf{MCC}           &  &  \\
\hline
Unsupervised                & 53.24          & 0.054          &  &  \\
Supervised                 & 57.36          & 0.128         &  &  \\
\textbf{Sup.\& Unsup.}          & \textbf{58.11}          & \textbf{0.153} &  & \\
\bottomrule
\end{tabular}
\end{table}

\section{Technical Experts}

To evaluate the performance of the number-to-text-alignment (prompt is shown in Table \ref{tab:recent_basic_stock_factors}), we carefully choose two popular time series baselines: 1) LLMTime~\cite{gruver2023large}: a single feature time series prediction method. Here we combine simply combine cross feature time series prediction sequences into one time series sequence; 2) LLMDate: following prompt in ~\cite{xie2023wall} which split time series features by date, we adopt similar prompt showing in appendix. We use the same base model and hyperparamters to evaluate the performance of the above prompts. As shown in Table \ref{tab:t9}, the performance of LLMTime is close to random, which indicates that the LLM is not good at cross feature time series prediction without meticulous task conversion. Besides, we find that N2I-Align performs the best, showing the effectiveness of number to text alignment.

\begin{table}[]
	\centering
	\caption{Comparison of the impact for stock movement prediction task over different technical analysis methods. }
	\label{tab:t9}
\begin{tabular}{lcc}
\bottomrule
\textbf{Method}                & \textbf{Accuracy}          & \textbf{MCC}        \\
\hline
Rand                & 50.68          & 0.002         \\
LLMTime                 & 51.12          & 0.028         \\
LLMDate                 & 53.47          & 0.082       \\
\textbf{N2I-Align}          & \textbf{59.62}          & \textbf{0.189} \\
\bottomrule
\end{tabular}
\end{table}

\section{Details of Verbalizer}
\label{sec:verb}
We define a verbalizer as an injective function that maps each label to a word list from vocabulary of large language model. For \emph{Up} label, we map it to a word list: [\emph{up}, \emph{boost}, \emph{positive}, \emph{rise}]. For \emph{Down} label, we map it to another word list: [\emph{down}, \emph{reduce}, \emph{negative}, \emph{drop}]

\section{Interpretablity Evaluation}
\subsection{Faithfulness} \label{sec:faithful}
Recent work has shown that LLMs are prone to suffer from hallucination generations across various applications, where the generated responses is either in conflict with existing fact or cannot be verified by the available knowledge resources. To evaluate how faithful the model response to the given knowledge, we design prompt  with help of GPT-4 to verify the faithfulness of model response (prompt is shown in Table \ref{tab:t21}). Unlike the conclusion in HaluEval that existing state of art LLM (GPT-3 at that time) can not distinguish between fact and hallucinated content, we find that GPT-4 is now capable of handle this job. To evaluate this statement, we samples 500 examples from ploutos response and invite  human labelers to annotate. For each query and model response, human labelers will annotate whether the response is faithful to the given knowledge resources (“Yes” or “No”) and list the corresponding reasons. Each response is labeled by three human labelers, and we adopt the max-voting strategy to determine the final faithfulness label. At last, we compare the Pearson correlation between the human labeled result and prediction from our proposed method, and achieved a Pearson correlation score of 0.826. The result indicates that this method can be used to evaluate the faithfulness of model response.

\begin{table*}[htp]  
\centering  
\caption{Technical analysis prompt example}  
\label{tab:recent_basic_stock_factors}  
  \resizebox{0.85\linewidth}{!}{%
\begin{tabular}{p{\linewidth}} 
\toprule  
  
\textbf{[Instruction]}:\\  
You are a seasoned stock market analyst expert in predicting future price trends based on historical stock factors.\\  
  
\textbf{[Stock Factors]}:\\  
From 2015-11-04 to 2015-11-08, some recent basic stock factors are presented below:\\
\textbf{Alpha: MV7 - Moving Average of 7 Days}\\  
Formula: ts\_mean(close, 7)\\  
Explanation: This Formula expression calculates the average closing price over the past 7 days, which helps to smooth out short-term volatility and identify the underlying trend in the price movement.\\  
Historical Values: 7798,7848,7878\\
\textbf{Alpha: MV20 - Moving Average of 20 Days}\\  
Formula: ts\_mean(close, 20)\\  
Explanation: Similar to MV7, this Formula expression computes the average closing price over a longer period of 20 days. This is often used to assess the medium-term market trend and\\ can act as support or resistance levels for the price.\\  
Historical Values: 7622,7644,7668\\
\textbf{Alpha: MACD - Moving Average Convergence Divergence}\\  
Formula: minus(ewma(close, 12), ewma(close, 26))\\  
Explanation: The MACD is a trend-following momentum indicator that shows the relationship between two moving averages of a security's price. The Formula expression represents the difference between the 12-day exponential moving average (EMA) and the 26-day EMA of the closing prices.\\  
Historical Values: -729,-747,-737\\
\textbf{Alpha: EMA - Exponential Moving Average}\\  
Formula: ewma(close, com=0.5)\\  
Explanation: This Formula expression calculates the exponential moving average of the closing prices, giving more weight to recent prices. The com parameter controls the degree of weighting decrease, making it a more sensitive measure of recent price movements.\\  
Historical Values: 8006,7931,7887\\
\textbf{Alpha: Bollinger Bands - Middle Line}\\  
Formula: ts\_stddev(close, 20)\\  
Explanation: This Formula expression is the simple moving average of the closing price over the past 20 days. It serves as the middle line in Bollinger Bands and is used to determine the intermediate-term trend.\\  
Historical Values: 202,206,202\\
\textbf{Alpha: Bollinger Bands - Upper Band}\\  
Formula: plus(ts\_mean(close, 20), times(ts\_stddev(close, 20), 2))\\  
Explanation: The upper band of the Bollinger Bands is calculated by adding two standard deviations to the 20-day moving average. This band adjusts for volatility and can signal overbought conditions when the price touches or breaches the upper band.\\  
Historical Values: 8027,8058,8073\\
\textbf{Alpha: Bollinger Bands - Lower Band}\\  
Formula: minus(ts\_mean(close, 20), times(ts\_stddev(close, 20), 2))\\  
Explanation: The lower band of the Bollinger Bands is calculated by subtracting two standard deviations from the 20-day moving average. It also adjusts for volatility and can signal oversold conditions when the price touches or breaches the lower band.\\  
Historical Values: 7217,7230,7263\\
\textbf{Alpha: LogMomentum}\\  
Formula: log(minus(close, shift(close, 1)))\\  
Explanation: This Formula expression calculates the natural logarithm of the difference between the current closing price and the previous day's closing price. It provides a measure of the momentum of the security's price by capturing the rate of change on a logarithmic scale.\\  
Historical Values: 437,436,434\\
\textbf{Alpha: VMA60 - Volume Moving Average 60}\\  
Formula: data['Volume'].rolling(60).mean() / (data['Volume'] + 1e-12)\\  
Explanation: The VMA60 Formula expression calculates the average volume over a 60-day period and divides it by the current volume plus a small constant to prevent division by zero. It compares the current volume to the average volume over a longer period, which can signal changes in trader participation.\\  
Historical Values: 113,131,129\\
\textbf{[Analysis]}:\\  
The historical price for XOM from 2015-11-05 to 2015-11-08 is 7895,7864. Please predict the XOM's price for the 2015-11-09 based on all above information.\\  
Answer:\\
  
\bottomrule  
\end{tabular}  
}
\end{table*}  

\begin{table*}[ht]
	\centering
	\caption{Unsupervised sentiment analysis based prompt}
	\label{tab:t2}
   \resizebox{0.85\linewidth}{!}{%
\begin{tabular}{p{\linewidth}} 
\toprule
\textbf{[Instruction]}: What is the sentiment of this news/tweet? \\
Please choose an answer from \{Negative/Neutral/Positive\}. \\
Input: Mr Ashley sold a 43pc stake in the company for more than pounds 900m at the time of the float. \\
\textbf{Answer}: \\
\bottomrule  
\end{tabular}}
\end{table*}

\section{PloutosGPT Training}
\subsection{Example for Rearview-mirror Instruction Prompt}
The prompt for rearview-mirror data scraping is shown in Table~\ref{tab:prompt_rearview}. 

\begin{table*}[htp]  
\centering  
\caption{Rearview-mirror instruction prompt example}  
\label{tab:prompt_rearview}  
  \resizebox{0.85\linewidth}{!}{%
\begin{tabular}{p{\linewidth}} 
\toprule
\textbf{[Instruction]}:\\  
You are a seasoned stock market analyst. Your task is to list the bullish or bearish rationales for companies based on relevant news and basic financials from the past weeks, then provide an analysis and prediction for the companies' stock price movement for the upcoming week. Your answer format should be as follow: \\

\textbf{Bullish Rationales}: \\  
  - Rationale 1 \\  
  - Rationale 2 \\  
  - ... \\  
\textbf{Bearish Rationales}: \\  
  - Rationale 1 \\  
  - Rationale 2 \\  
  - ... \\  
\textbf{Prediction}: \\  
  - Your prediction here in less than 100 words. \\  

\textbf{[Sentiment Expert Input]:}\\
From 2015-07-29 to 2015-08-02, Company news during this period are listed below:\\
... (news used in sentiment analysis prompt as shown in Table \ref{tab:t3}).\\
\textbf{[Sentiment Expert Prediction]:} Down\\
\textbf{[Technical Expert Input]:}\\
From 2015-07-29 to 2015-08-02, Company news during this period are listed below:\\
... (alphas used in technical analysis prompt as shown in Table \ref{tab:recent_basic_stock_factors}).\\
\textbf{[Technical Expert Prediction]:} Up\\
\textbf{[Analysis]}: \\  
Assume your prediction for XOM in 2015-08-03 is down by 1-2\%, generate the rationales based on all the information before 2015-08-03 to justifies your prediction, ensuring that the prediction itself does not serve as a basis for the analysis but as a reasoned outcome of it. Please note that the generated rationale should present in or can be inferred from any one given information or combination of multiple given information. Then, conclude with a short prediction (less than 100 words) to decide the price movement prediction.\\

\bottomrule  
\end{tabular}  
}
\end{table*}

\begin{table*}[ht]
	\centering
	\caption{Prompt for faithfulness measurement}
	\label{tab:t21}
   \resizebox{0.85\linewidth}{!}{%
\begin{tabular}{p{\linewidth}} 
\toprule
\textbf{[Instruction]}: You will be given a yes/no question regarding if a rationale is either present or can be inferred from the given information. Your task is to answer yes or no and tell me why based on given information given and guidelines below. \\
Answer yes if:\\
The rationale is present in any one given information or combination of multiple given information.\\
The rationale can be inferred from any one given information or combination of given information.\\
The rationale is more general than the rationales in the given information, and it can be obtained by simplifying the rationales in the given information.\\
If rationale in question is supported by at least one given information or follows any of the above or a combination of above conditions answer will be yes, even if other given information contradict it.\\
Answer no if:\\
The rationale is not present in any of the given information individually or combined, or it cannot be clearly inferred from any of the given information individually or combined.\\
The rationale is more specific than the rationales in any of the given information.\\
The rationale is contradictory to the rationales in all the given information i.e. there is no given information which supports the rationale.\\
If rationale in question contains some topic which are neither present nor inferred from any of the given information, then answer no to the question.\\

\textbf{[Knowledge]}: ...\\
\textbf{[Rationale]}: ...\\
\textbf{Answer}: \\
\bottomrule  
\end{tabular}}
\end{table*}

\end{document}